\documentclass[12pt]{article}

\usepackage{mathtools}
\usepackage{color}
\usepackage{amsmath}
\usepackage{graphicx}
\usepackage{amssymb}
\usepackage{amsthm}
\usepackage{upgreek}

\newcommand{\omegabold}{\boldsymbol\omega}

\title{Toward analytic theory of the Rayleigh--Taylor instability: \\lessons from a toy model}

\author{Alexei A. Mailybaev\footnote{Instituto Nacional de Matem\'atica Pura e Aplicada -- IMPA, Rio de Janeiro, Brazil. E-mail: alexei@impa.br}
}

\begin{document}

\maketitle

\begin{abstract}
In this work we suggest that a turbulent phase of the Rayleigh--Taylor instability can be explained as a universal stochastic wave traveling with constant speed in a properly renormalized system. This wave, originating from ordinary deterministic chaos in a renormalized time, has two constant limiting states at both sides. These states are related to the initial discontinuity at large scales and to stationary turbulence at small scales. The theoretical analysis is confirmed with extensive numerical simulations made for a new shell model, which features basic properties of the phenomenological theory for the Rayleigh--Taylor instability.
\end{abstract}

\section{Introduction}
\label{sec1}

The Rayleigh--Taylor (RT) instability arises at an interface between two fluids of different densities in the presence of gravity or acceleration. Linear analysis for ideal fluid predicts that a small perturbation with wavenumber $k$ grows exponentially with the eigenvalue $\lambda \sim \sqrt{k}$, which means an explosive growth at small scales~\cite{kull1991theory}. Propagation of a disturbance to larger and larger scales due to nonlinear interaction generates a growing mixing layer with turbulent dynamics. Occurrence of the RT instability is abundant in nature, which includes astrophysical, geological and atmospheric phenomena, as well as various technological applications such as  combustion. We refer to~\cite{abarzhi2010review,andrews2010small,boffetta2016incompressible} for recent reviews describing theoretical, numerical and experimental advances in this area.

Phenomenological theory for a nonlinear stage of the RT instability in the Boussinesq equations was developed in~\cite{chertkov2003phenomenology}. It largely relies on arguments of the classical Kolmogorov theory of turbulent flow~\cite{frisch1995turbulence}, extended to account for gravity and non-stationarity. This approach provides statistical predictions for the growth of a mixing layer and scaling of velocity and temperature fluctuations in the inertial interval and dissipative scales. Qualitatively different theories follow for two (2D) and three (3D) spatial dimensions. The phenomenological predictions were shown to be in reasonable agreement with numerical simulations~\cite{cabot2006reynolds,boffetta2009kolmogorov,zhou2013temporal}. However, measurable deviations were observed for scaling exponents~\cite{celani2006rayleigh,biferale2010high}, leading to anomalous corrections in direct analogy with the hydrodynamic turbulence. The full theoretical understanding of the RT instability remains a big challenge in fluid mechanics. It includes a description of small-scale behavior, which may (or may not) be equivalent to the developed isotropic turbulence, as well as strongly anisotropic and non-stationary dynamics at large scales.

In this paper, we propose the analytic approach that goes beyond the phenomenological theory. We suggest that with a proper renormalization one maps the RT dynamics into a relatively simple object -- a stochastic traveling wave, which can be understood as the consequence of ordinary deterministic chaos in renormalized time. The basic idea is inspired by the explanation of spontaneously stochastic solutions developing from a blowup state in inviscid shell models of turbulence~\cite{mailybaev2016spontaneous,mailybaev2015spontaneously}. Here, one expands the evolution to a semi-infinite interval $(-\infty,\, \tau]$ using a logarithmic time variable $\tau = \log_h t$. Thereby, the solution is determined by a probability measure of a chaotic attractor in the new system. 

For numerical analysis, we create a new shell model for the RT instability. This model is based on a discrete number of scales, $r_n = h^{-n}$ with $n = 1,2,\ldots$ and $h > 1$, and it mimics all basic properties underlying the phenomenological theory in~\cite{chertkov2003phenomenology}. With a large number of accurate simulations ($10^5$ independent simulations with random initial perturbations for each case), a convincing confirmation of the proposed theoretical construction is given. A stochastic RT wave traveling from small to large scales at a constant speed is clearly observed in renormalized variables, separating two constant limiting states corresponding to the initial temperature jump and stationary turbulence. The RT wave occupies two and three decades of spatial scales for the 2D and 3D shell models, respectively, suggesting that a similar numerical analysis is feasible for the full Boussinesq system.

The paper is organized as follows. We start with basic facts of the RT instability in Section~\ref{sec2}. Section~\ref{sec3} introduces a new shell model, which describes the RT instability as confirmed in Section~\ref{sec3b}. Section~\ref{sec4} describes a renormalization scheme. Sections~\ref{sec5} and \ref{sec6} analyze the shell models that describe the RT instability in two and three dimensions, respectively. We end with some conclusions.

\section{Basics of the Rayleigh--Taylor instability}
\label{sec2}

Let us consider an incompressible buoyancy-driven flow in unbounded space $\mathbf{r} = (x,y,z) \in \mathbb{R}^3$ or plane $\mathbf{r} = (x,z) \in \mathbb{R}^2$. In Boussinesq approximation, the flow is governed by the equations~\cite{landau1987fluid}
\begin{equation}
\partial_t \mathbf{u} + \mathbf{u}\cdot \nabla \mathbf{u} 
= -\nabla p+\nu \nabla^2 \mathbf{u}+\beta g\mathbf{e}_zT,
\label{eq1}
\end{equation}
\begin{equation}
\partial_t T + \mathbf{u}\cdot \nabla T
= \kappa \nabla^2 T,
\label{eq2}
\end{equation}
\begin{equation}
\nabla \cdot \mathbf{u} = 0,
\label{eq3}
\end{equation}
where $\mathbf{u}\in \mathbb{R}^3$ (or $\mathbb{R}^2$) is the velocity, $T \in \mathbb{R}$ is the temperature and $\mathbf{e}_z = (0,0,1)$ is the unit vector in vertical direction. The constant parameters are the viscosity $\nu$, the thermal conductivity $\kappa$ and the product $\beta g$ of the thermal expansion coefficient with the gravitational acceleration. In this description, a warmer fluid is assumed to be lighter than a colder fluid. The Rayleigh--Taylor (RT) instability refers to the initial condition 
\begin{equation}
t = 0:\quad
\mathbf{u} = 0,\quad
T = -\sigma \Theta\,\mathrm{sgn}\,z,
\label{eq4}
\end{equation}
where $\sigma = \pm 1$ and $\Theta > 0$ is half of the temperature jump. This initial condition describes the fluid composed of two layers with different temperatures in the upper and lower half-spaces. 

In the ideal fluid, $\nu = \kappa = 0$, the initial state (\ref{eq4}) is an equilibrium. Then the relation~\cite{kull1991theory,boffetta2016incompressible 
\begin{equation}
\lambda = \pm\sqrt{(\sigma\Theta \beta g) k}
\label{eq5}
\end{equation}
}defines the growth rate $\propto e^{\lambda t}$ for a small single-mode perturbation of the interface $z = 0$ with the wavenumber $k$. The configuration with $\sigma = -1$, when the warmer fluid is above the colder fluid, is stable because the exponent $\lambda \propto \pm i\sqrt{k}$ is purely imaginary. The Rayleigh--Taylor instability occurs in the opposite case of $\sigma = 1$, when the warmer fluid is below. In this case the real exponents $\lambda \propto \sqrt{k}$ are positive and unbounded for large $k$, which corresponds to explosive growth of small-scale perturbations. 

\begin{figure}
\centering
\includegraphics[width=0.65\textwidth]{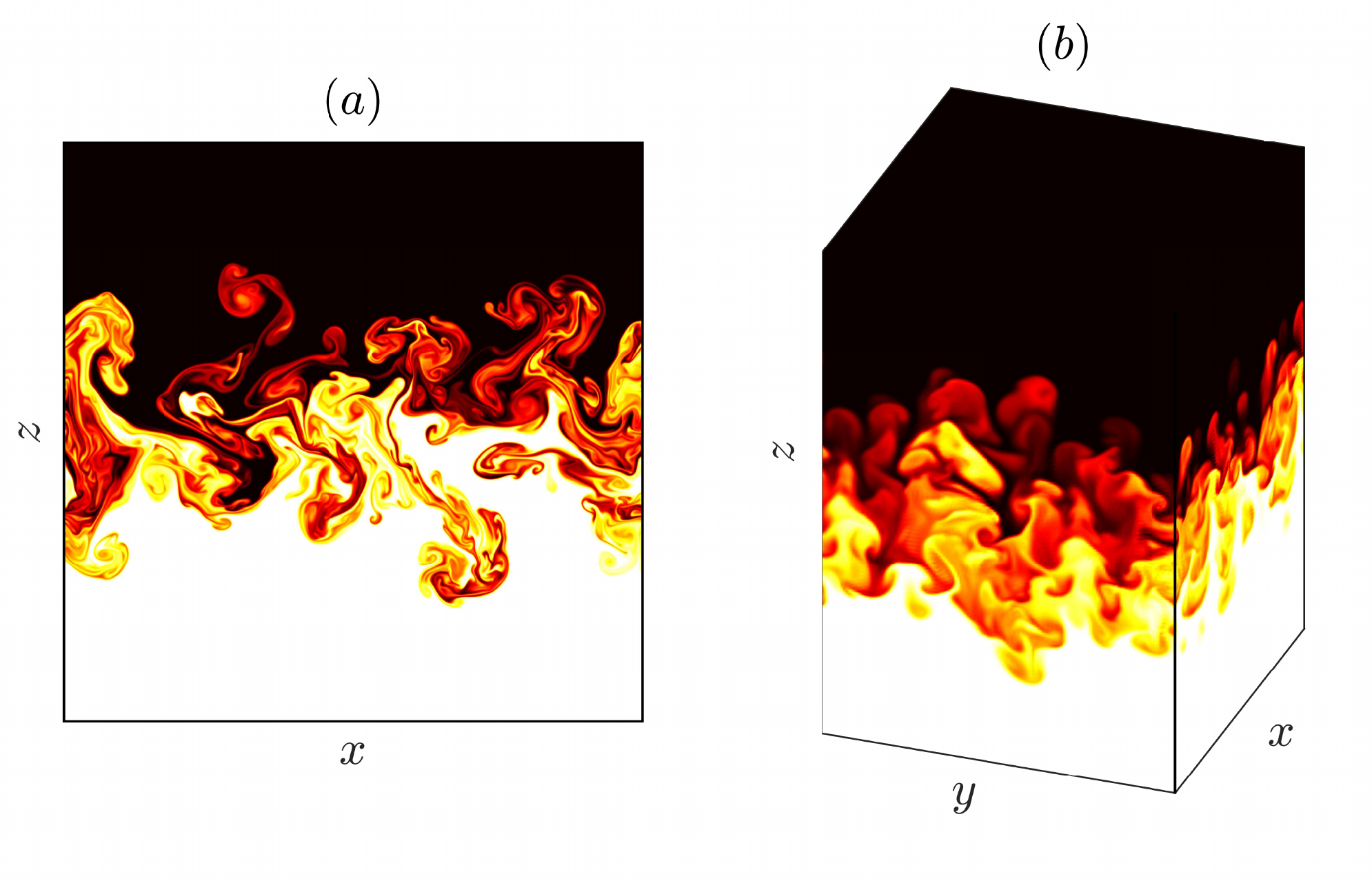}
\caption{(a) 2D and (b) 3D Rayleigh--Taylor instability in a periodic domain, developing from a small-scale random perturbation of initial temperature jump (\ref{eq4}) with $\sigma = 1$. Results are obtained by direct numerical simulations of system (\ref{eq1})--(\ref{eq3})  for small dissipation coefficients using the pseudo-spectral method. The warm (white) fluid is below the cold (black) fluid.}
\label{fig1}
\end{figure}

In the unstable configuration ($\sigma = 1$), a small perturbation after a rapid linear stage develops into a strongly nonlinear flow, involving larger and larger scales with increasing time, Fig.~\ref{fig1}. This generates turbulent dynamics  in a layer of characteristic width $L(t)$ and velocity $u_L(t)$ around the initial interface $z = 0$. When this layer gets large, diffusive effects become negligible at scale $L(t)$. With the remaining dimensional parameters $\beta g$ [m/s$^2$/K] and $\Theta$ [K], the dimensional prediction for asymptotic growth of the mixing layer can be written in the unique way as
\begin{equation}
L(t) \sim (\beta g \Theta)\,t^2, \quad u_L(t) \sim (\beta g \Theta)\,t.
\label{eq6}
\end{equation}
Viscosity is important at the much smaller Kolmogorov viscous scale $\eta(t)$ specified below, and we denote by $r_d(t)$ an analogous small scale for thermal conduction. Thus, the so-called inertial interval $L(t) \gg r \gg \max(\eta(t),r_d(t))$ is formed, which separates the integral scale from the dissipative ones. 

In this work, we focus on the dynamics at large and inertial-interval scales, thus, leaving aside the dissipative effects at smaller scales. According to the phenomenological theory, that we describe here following~\cite{chertkov2003phenomenology}, development of the RT instability is qualitatively different in two- and three-dimensional spaces. In three dimensions, the inertial interval is dominated by a nonlinear transfer of kinetic energy from large to small scales, while the buoyancy term is negligible. The mean energy flux to small scales can be estimated as $\varepsilon(t) \sim u_L^3/L$. With the quasi-stationarity assumption, this energy flux can be written as $\varepsilon(t) \sim \delta u_r^3/r$ for velocity fluctuations $\delta u_r$ at any scale $r$ in the inertial interval. Similarly, passively advected temperature fluctuations develop the flux $\varepsilon_T(t) \sim \delta T_r^2\delta u_r/r \sim \Theta^2 u_L/L$. This yields the Kolmogorov scaling
\begin{equation}
\mathrm{3D}:\quad
\delta u_r(t) \sim u_L(t) \left(\frac{r}{L(t)}\right)^{1/3},\quad 
\delta T_r(t) \sim  \Theta\left(\frac{r}{L(t)}\right)^{1/3}.
\label{eq7}
\end{equation}
One can check with relations (\ref{eq6}) and (\ref{eq7}) that $\beta g\,\delta T_r \ll \delta u_r^2/r$ for $r \ll L$, justifying the hypothesis that the buoyancy term is negligible in Eq.~(\ref{eq1}) at small scales. It must be emphasized however, that the dynamics in the inertial range is intermittent~\cite{boffetta2016incompressible}, which implies anomalous corrections for the exponents in relations like (\ref{eq7}).

The viscous scale $\eta(t)$ is estimated by comparing the nonlinear term $\delta u_r^2/r$ with the viscous term $\nu\delta u_r/r^2$ at $r \sim \eta$. Using (\ref{eq6}) and (\ref{eq7}), one obtains 
\begin{equation}
\mathrm{3D}:\quad
\eta(t) \sim  (\beta g \Theta)^{-1/2}\nu^{3/4}t^{-1/4},
\label{eq7b}
\end{equation}
showing that the viscous scale decreases with time.
When $\nu \sim \kappa$, thermal dissipation becomes important at the same small scale $r_d(t) \sim \eta(t)$.  For $\nu \gg \kappa$, further analysis~\cite{chertkov2003phenomenology} provides $r_d(t) \sim \eta(t)\sqrt{\kappa/\nu} \ll \eta(t)$. Condition $\eta \ll L$ with relations (\ref{eq6}) and (\ref{eq7b}), yield the lower bound for the time $t \gg (\beta g \Theta)^{-2/3}\nu^{1/3}$ and width $L \gg (\beta g \Theta)^{-1/3}\nu^{2/3}$ that allow for existence of the inertial interval.

Quite a different phenomenology corresponds to the two-dimensional RT instability, when $\mathbf{r} = (x,z) \in \mathbb{R}^2$. In this case, the cascade of kinetic energy to small scales is not possible due to the enstrophy, which is a second inviscid invariant in a 2D flow. As a result, the buoyancy term is not negligible and leads to the so-called Bolgiano--Obukhov scenario. This scenario is characterized by the cascade of temperature fluctuations to small scales with the flux $\varepsilon_T(t) \sim \delta T_r^2\delta u_r/r \sim \Theta^2u_L/L$. Additionally, the buoyancy term $\beta g \delta T_r$ matches the nonlinear term $\delta u_r^2/r$ in Eq.~(\ref{eq1}). Using (\ref{eq6}), this provides the relations 
\begin{equation}
\mathrm{2D}:\quad
\delta u_r(t) \sim u_L(t) \left(\frac{r}{L(t)}\right)^{3/5},\quad 
\delta T_r(t) \sim \Theta\left(\frac{r}{L(t)}\right)^{1/5}.
\label{eq8}
\end{equation}
Estimating the viscous scale as in the 3D case yields
\begin{equation}
\mathrm{2D}:\quad
\eta(t) \sim (\beta g \Theta)^{-1/4}\nu^{5/8}t^{1/8}. 
\label{eq8b}
\end{equation}
Contrary to the 3D case, this scale grows with time.
From the condition $\eta \ll L $ we get the same bounds $t \gg (\beta g \Theta)^{-2/3}\nu^{1/3}$ and $L \gg (\beta g \Theta)^{-1/3}\nu^{2/3}$ compatible with the existence of inertial interval.
Again, one may expect anomalous corrections due to intermittency: such corrections for temperature fluctuations were confirmed by numerical simulations in~\cite{biferale2010high}, though the question of intermittency for velocity statistics remains open.

\section{Shell model of the Rayleigh--Taylor instability}
\label{sec3}

In this section, we create a ``toy model'' that possesses all properties of the RT instability described in Section~\ref{sec2}. We will construct this model based on a geometric progression of discrete scales $r_n = h^{-n}$ with $h > 1$ and $n = 1,2,\ldots$.  Note that the scaling symmetry is a key feature underlying the RT phenomenology, and a geometric progression is the simplest possible scaling-invariant representation, where the shift $n \mapsto n+1$ stands for $r_n \mapsto r_{n+1} = r_n/h$. At each scale $r_n$ (also called shell), we represent the velocity fluctuations by a real number $u_n \in \mathbb{R}$ and associate $\omega_n = u_n/r_n = k_nu_n$ with the vorticity fluctuations, where $k_n = 1/r_n = h^n$ is the wavenumber. For the temperature field, we have to distinguish  horizontal temperature fluctuations $R_n \in \mathbb{R}$ and vertical temperature fluctuations $T_n \in \mathbb{R}$. Shell models of this kind represent a common tool for testing theoretical ideas on statistical behavior in developed turbulence~\cite{biferale2003shell}.

Equations of motion are formed similarly to the Obukhov and Desnyansky--Novikov models~\cite{obukhov1971some, desnyansky1974evolution}, i.e., limiting interactions to the neighboring shells; see also \cite{brandenburg1992energy,Ching2010,mailybaev2012c} for shell models of natural convection. First, it is convenient to rewrite Eq.~(\ref{eq1}) in terms of vorticity as
\begin{equation}
\frac{\partial\omegabold}{\partial t} - \nu \nabla^2\omegabold
= \mathrm{rot}\,(\mathbf{v}\times \omegabold)
+\beta g \left(\partial_yT\mathbf{e}_x-\partial_xT\mathbf{e}_y\right),\quad
\mathbf{v} = \mathrm{rot}^{-1}\omegabold,
\label{eq9}
\end{equation}
where the buoyancy terms contain only horizontal derivatives of temperature.
Then the shell model we propose reads
\begin{equation}
\dot{\omega}_n+\nu k_n^2 \omega_n 
= \left[\omega_{n-1}^2-c\omega_n\omega_{n+1}+b(\omega_{n-1}\omega_n-c\omega_{n+1}^2)\right]
+k_nR_n,
\label{eq10}
\end{equation}
\begin{equation}
\dot{R}_n +\kappa k_n^2 R_n = \omega_nR_{n+1}-\omega_{n-1}R_{n-1}+\gamma \omega_nT_n,
\label{eq11}
\end{equation}
\begin{equation}
\dot{T}_n +\kappa k_n^2 T_n = \omega_nT_{n+1}-\omega_{n-1}T_{n-1}-\gamma \omega_nR_n,
\label{eq12}
\end{equation}
where $c$, $b$ and $\gamma > 0$ are real parameters specified later, and dots denote derivatives with respect to time. A large-scale boundary condition is chosen as $\omega_0 = R_0 = T_0 = 0$.
One can see that Eqs.~(\ref{eq10})--(\ref{eq12}) contain the terms that mimic viscous terms (on the left-hand side) and nonlinear together with buoyancy terms (on the right-hand sides) of the original Eqs.~(\ref{eq9}) and (\ref{eq2}). Following (\ref{eq9}), only the variables $R_n$ corresponding to horizontal temperature fluctuations appear in (\ref{eq10}). The last terms in (\ref{eq11}) and (\ref{eq12}) model the transition between horizontal and vertical  temperature fluctuations due to rotation. Equations (\ref{eq11}) and (\ref{eq12}) are designed to have the inviscid invariant $S = \sum(R_n^2+T_n^2)$ measuring the temperature fluctuations (entropy). 

A choice of the coefficient $c$ allows to form an extra inviscid invariant, which is a crucial point for our model. As we mentioned in Section~\ref{sec2}, in three dimensions, the inertial interval is dominated by a nonlinear transfer of kinetic energy from large to small scales, while the buoyancy term is negligible. This requires to have the kinetic energy as the inviscid invariant for the system with no buoyancy term. In our shell model, this is achieved by taking $c  = 1/h^2$ and defining the kinetic energy as $E = \sum u_n^2 = \sum \omega_n^2/k_n^2$. Similarly, in two dimensions, the Bolgiano--Obukhov dynamics is dominated by the enstrophy, which is conserved in the flow with no buoyancy term. In the shell model, this property is ensured by choosing $c = 1$ with the enstrophy defined as $\Omega = \sum \omega_n^2$. We summarize this classification as
\begin{equation}
\textrm{3D}:\quad c  = 1/h^2;\quad
\textrm{2D}:\quad c  = 1.
\label{eq13}
\end{equation}
As we will see below, the resulting shell model successfully demonstrates basic properties of the RT instability, confirming once again that the right choice of invariants was made. It should be noted that conservation of the total (kinetic plus potential) energy as an extra inviscid invariant (as well as some other propertied of the flow) could be achieved by considering more sophisticated shell models, but this analysis goes beyond the scope of this paper. As for the parameters $b$ and $\gamma$ in (\ref{eq10})--(\ref{eq12}), their specific values are not so important, but have to be chosen such that the model has a chaotic dynamics.

For the RT instability, we choose the initial condition
\begin{equation}
t = 0:\quad \omega_n  = 0;\quad
R_n = 0,\quad T_n = \sigma,\quad n = 1,2,3,\ldots,
\label{eq14}
\end{equation}
that mimics a jump with $\sigma\Theta = \pm 1$ for vertical temperature fluctuation $T_n$. This initial condition is clearly an equilibrium of our shell model in the absence of dissipative terms, $\nu = \kappa = 0$. Note that the dimensional parameters corresponding to $\beta g$ in (\ref{eq1}) and $\Theta$ in (\ref{eq4}) are both chosen as unity in the shell model.

We should stress that introducing two types of temperature variables in the shell model is fundamental, as it follows from the physics of RT instability: The system must be in equilibrium for an arbitrary vertical temperature distribution, when the temperature is constant in horizontal direction. In this setting, only horizontal changes of temperature  introduce the torque necessary for vorticity generation. In turn, this vorticity rotates the temperature field amplifying its horizontal fluctuations. All these properties are preserved in the proposed shell model, where the variables $R_n$ and $T_n$ can be interpreted, for example, as horizontal and vertical Fourier components of the temperature field.

\section{Linear analysis and phenomenology of turbulent mixing in a shell model}
\label{sec3b}

Inviscid equations (\ref{eq10})--(\ref{eq12}) linearized near equilibrium (\ref{eq14}) take the form
\begin{equation}
\Delta\dot{\omega}_n
= k_n \Delta R_n,
\quad
\Delta\dot{R}_n = \sigma\gamma\Delta \omega_n,
\quad
\Delta\dot{T}_n = \sigma(\Delta\omega_{n}-\Delta\omega_{n-1}).
\label{eq15}
\end{equation}
Considering the time dependence proportional to $e^{\lambda t}$, the first two equations yield the spectrum 
\begin{equation}
\lambda = \pm\sqrt{\sigma  \gamma k_n},
\label{eq18}
\end{equation}
which has the same form as (\ref{eq5}). Additionally, the system possesses an infinite number of neutral modes with $\lambda = 0$, since any vertical distribution of the temperatures $T_n$ with $\omega_n = R_n = 0$ is an equilibrium, similarly to the original continuous system. We see that the equilibrium state (\ref{eq14}) is stable for $\sigma = -1$ and unstable for $\sigma = 1$. The latter case is attributed to the RT instability in our shell model.

The phenomenological theory of RT instability can be deduced for the shell model (\ref{eq10})--(\ref{eq12}) following just the same arguments as for the Boussinesq equations in Section~\ref{sec2}. A small generic perturbation develops rapidly at small scales $r_n = h^{-n}$ (large $n$),  because the linear instability is dominated by larger exponents $\lambda = \sqrt{k_n}$ for larger $k_n = h^n$. Then oscillations propagate to larger and larger scales (smaller $n$) due to nonlinear interaction. Thus, in terms of shell numbers, the disturbance propagates from large to small $n$. The mixing layer width $L(t)$ can be related to the characteristic shell number $N(t)$ reached by the RT instability at time $t$ as $L(t) = r_N = h^{-N(t)}$. 

Equations~(\ref{eq6}) of the phenomenological theory are written in terms of shell variables as
\begin{equation}
L(t) = h^{-N(t)} \sim  t^2, \quad u_N(t) \sim  t,
\label{eq19}
\end{equation}
because $\beta g\Theta = 1$ in the shell model.
Similarly, the power laws (\ref{eq7}) and (\ref{eq8}) for scales in the inertial interval, $\max(\eta(t),r_d(t)) \ll r_n \ll L(t)$, are written as
\begin{equation}
\mathrm{3D}:\quad
u_n(t) \sim u_N(t) \left(\frac{r_n}{L(t)}\right)^{1/3},\quad 
R_n(t) \sim T_n(t) \sim  \left(\frac{r_n}{L(t)}\right)^{1/3}.
\label{eq20}
\end{equation}
\begin{equation}
\mathrm{2D}:\quad
u_n(t) \sim u_N(t) \left(\frac{r_n}{L(t)}\right)^{3/5},\quad 
R_n(t) \sim T_n(t) \sim  \left(\frac{r_n}{L(t)}\right)^{1/5}.
\label{eq21}
\end{equation}
The Kolmogorov viscous scales are given by the same Eqs.~(\ref{eq7b}) and (\ref{eq8b}).

For comparison with numerical simulations of the shell model, we integrate Eqs.~(\ref{eq10})--(\ref{eq12}) with $n = 1,\ldots,40$ total shells and the parameters $h = 2$, $\nu = \kappa = 10^{-14}$, $b = 0.1$. The remaining parameter is taken as $\gamma = 1$ for the 2D model and $\gamma = 0.7$ for the 3D model. A tiny random initial perturbation is given to the horizontal temperature variable $R_n$ at the shell $n = 29$, slightly above the scale $r_n \sim \nu^{2/3}$, see Section~\ref{sec2}. For the statistical analysis, $10^5$ independent simulations were performed for each model. 

\begin{figure}
\centering
\includegraphics[width=0.75\textwidth]{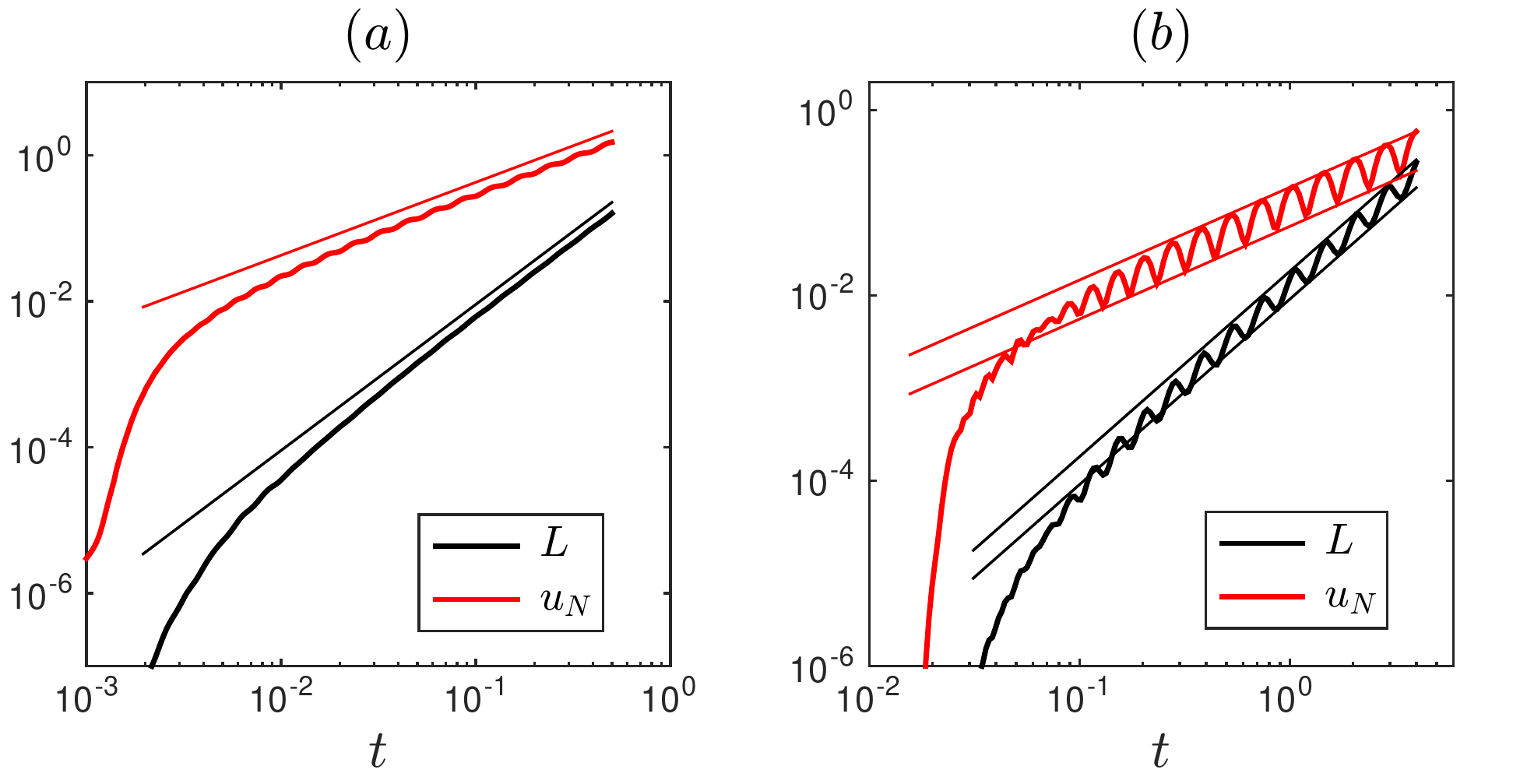}
\caption{Evolution of the mixing-layer width $L(t)$ and large-scale velocity $u_N(t)$ in logarithmic scale for (a) 2D and (b) 3D model. Thin straight lines show the slopes for power laws $L \sim t^2$ and $u_N \sim t$. In the right figure, two straight lines are plotted for each curve to show periodicity at large times.}
\label{fig2}
\end{figure}

In agreement with the linear analysis of ideal model in (\ref{eq18}), the initial conditions with $\sigma = -1$ do not lead to instability, demonstrating only a slow increase of the viscous range. This case corresponds to the stable configuration with a warmer fluid on the top. On the contrary, for $\sigma = 1$, after a very fast linear growth, the solution develops chaotic oscillations at small scales, which propagate to larger and larger scales (smaller shell numbers). Let us define the mixing layer width and the large-scale velocity as
\begin{equation}
L(t) = \sum_n \langle 1-T_n(t)\rangle r_n,\quad
u_N(t) = \Big\langle\sum_n u_n^2(t) \Big\rangle^{1/2}
\label{eq22}
\end{equation}
where the averaging is made at fixed time $t$ over an ensamble of $10^5$ independent simulations. These expressions are analogous to the integral definition used for the continuous model, see e.g. \cite{biferale2010high}. Figure~\ref{fig2} shows the numerical results confirming the dimensional prediction (\ref{eq19}). Note that a small periodic oscillation around the power-law average value is an artifact of the shell model, which contains only discrete scales $r_n$. Indeed, unlike in the continuous model, the ``mixing layer'' in a shell model can only reproduce itself at discrete times, when it grows from scale $r_n$ to $r_{n-1}$. These times are proportional to $t_n \sim r_n^{1/2}$, leading to  periodic oscillations in logarithmic scale of Fig.~\ref{fig2}.

\begin{figure}[t]
\centering
\includegraphics[width=0.75\textwidth]{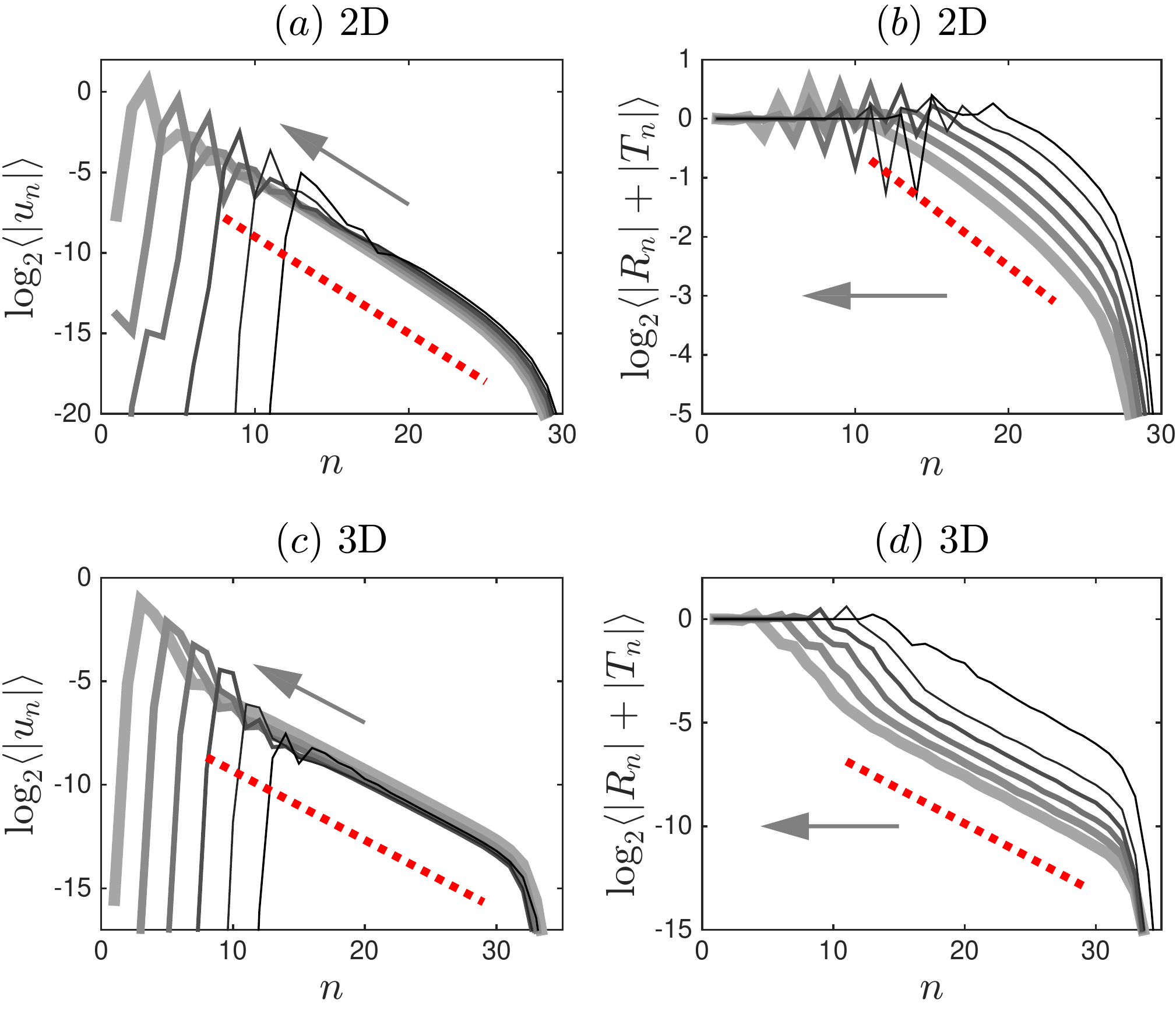}
\caption{Logarithms of the first moments for shell velocities, $u_n$, and temperatures $R_n$, $T_n$ as functions of the shell number $n$. Solid (black to grey) lines represent the result of averaging over $10^5$ random perturbations of initial conditions at fixed times $t = 2^{-6},2^{-5},\ldots,2^{-1}$ for the 2D model and at $t = 2^{-3},2^{-2},\ldots,2^2$ for the 3D model; thicker lines are used for larger times. Arrows indicate the dynamics with increasing time. Dotted red lines show theoretical slopes (\ref{eq20}) and (\ref{eq21}) at small scales.}
\label{fig_new}
\end{figure}

For verification of the predictions of phenomenological theory, we plot in Fig.~\ref{fig_new} the first moments of shell velocities and temperatures, which are averaged over $10^5$ independent simulations at fixed times. Numerical results agree very well with theoretical slopes of the Kolmogorov (\ref{eq20}) and Bolgiano-Obukhov (\ref{eq21}) scenario demonstrated by the dotted red lines. We will show in Section~\ref{sec5.2} that temperatures in the 2D model develop anomalous corrections.

\section{Stochastic traveling wave in renormalized system}
\label{sec4}

For understanding a detailed mechanism of the RT instability, we propose the renormalized form of model equations. Let us introduce the new (logarithmic) time variable $\tau$ and new dependent variables denoted with tildes as
\begin{equation}
t = h^{\tau},\quad 
\omega_n = h^{-\tau}\tilde{\omega}_n,\quad
R_n = h^{-n-2\tau}\tilde{R}_n,\quad
T_n = h^{-n-2\tau}\tilde{T}_n.
\label{eq23}
\end{equation}
Since we are interested in the dynamics at integral and inertial interval scales, we will drop the dissipative terms in our analysis. Then, in the new variables, inviscid Eqs.~(\ref{eq10})--(\ref{eq12}) take the form
\begin{equation}
\frac{1}{\log h}\frac{d\tilde \omega_n}{d\tau}
= \tilde \omega_n+\tilde \omega_{n-1}^2-c\tilde \omega_n\tilde \omega_{n+1}
+b(\tilde \omega_{n-1}\tilde \omega_n-c\tilde \omega_{n+1}^2)+ \tilde R_n,
\label{eq24}
\end{equation}
\begin{equation}
\frac{1}{\log h}\frac{d\tilde R_n}{d\tau}
 = 2\tilde R_n+h^{-1}\tilde \omega_n\tilde R_{n+1}-h\tilde \omega_{n-1}\tilde R_{n-1}+\gamma\tilde \omega_n\tilde T_n,
\label{eq25}
\end{equation}
\begin{equation}
\frac{1}{\log h}\frac{d\tilde T_n}{d\tau}
 = 2\tilde T_n+h^{-1}\tilde \omega_n\tilde T_{n+1}-h\tilde \omega_{n-1}\tilde T_{n-1}-\gamma\tilde \omega_n\tilde R_n.
\label{eq26}
\end{equation}

It is remarkable that the new system (\ref{eq24})--(\ref{eq26}) is translation invariant both in the new time $\tau$ and in the shell number $n$, which simply reflects the scaling invariance of the original model. Another key property is that the initial time $t = 0$ corresponds $\tau \to -\infty$, i.e., the relevant solution of the renormalized inviscid system is the one corresponding to an infinitely long evolution, i.e., an attractor. The translation invariance allows an attractor to be a traveling wave. For example, such solutions can be steady-state waves traveling with a constant speed $v$, i.e., depending only on a single variable $\xi = n-v\tau$. Alternatively, this can be a wave with a pulsating (periodically or chaotically) state moving with an average speed $v$, see \cite{mailybaev2016spontaneous,mailybaev2015spontaneously} for some examples. These types of waves can be seen as direct analogs of fixed-point, periodic or chaotic attractors in dynamical systems. 
Assuming a traveling wave solution, the speed can be found immediately as $v = -2$ by comparing (\ref{eq14}) with $\sigma = 1$ and (\ref{eq23}), which yields
\begin{equation}
\tilde T_n \to h^{n+2\tau}\quad \textrm{as}\quad \tau \to -\infty \ \ \textrm{(fixed }n\textrm{)}.
\label{eq27}
\end{equation}
Negative sign of the speed implies that the wave moves from large to small shell numbers $n$ (from small to large scales $r_n$).
We will clearly demonstrate in the following sections that the RT instability in our model is described by a chaotic wave traveling with average speed $v = -2$, both in the 2D and 3D cases.

For interpretation of the results, it is useful to discuss some implications of a chaotic wave in the renormalized model. Due to exponential separation of trajectories in a chaotic system, we expect that the information on the initial state is rapidly forgotten at times $\sim \tau_*$ corresponding to the transient from an initial state to a chaotic attractor. At later times, physical description of the dynamics is given by the chaotic attractor, i.e., the relevant physical solution is a stochastic process (an invariant probability measure) moving as a traveling wave from larger to smaller shell numbers. This description becomes exact in the inviscid limit, when both dissipation parameters $\nu,\kappa \to 0$: in this limit the initialization process is moved to $\tau_* \to -\infty$ corresponding to a vanishing transient time $t_* = h^{\tau_*} \to 0$. This, in particular, implies that the solution becomes stochastic immediately for $t > 0$, i.e., the RT instability is an example of the spontaneous stochasticity phenomenon. Another key observation is that the resulting stochastic solution is unique (universal) provided that there is a unique chaotic attractor for the renormalized system.

The next feature that is substantial for representating the RT instability as a stochastic traveling wave refers to the so-called third Kolmogorov hypothesis. This hypothesis suggests that the ratios of velocity increments have universal probability distribution at small scales~\cite{kolmogorov1962refinement}, which was verified extensively for shell models~\cite{benzi1993intermittency,eyink2003gibbsian} and the Navier--Stokes equations~\cite{chen2003kolmogorov}. There is a large freedom of choosing such ratios in our model. For example, it is convenient to use the ratios (multipliers)
\begin{equation}
\rho_n^{\omega} = \frac{\tilde \omega_n}{\tilde \omega_{n-1}} 
= \frac{\omega_n}{\omega_{n-1}} ,\quad
\rho_n^{R} = \frac{\tilde R_n}{\tilde R_{n-1}}
= \frac{h R_n}{R_{n-1}},\quad
\rho_n^{T} = \frac{\tilde T_n}{\tilde T_{n-1}}
= \frac{h T_n}{T_{n-1}},
\label{eq28}
\end{equation}
where we also provided the expressions in terms of the original variables (with no tildes).
Note that the statistic description in terms of multipliers (\ref{eq28}) is complete, because there is a one-to-one relation between the multipliers and the original variables, except for a zero-measure set when any of the original variables vanishes. Of course, other ratios can also be used for the same purpose.

\section{RT instability in 2D case}
\label{sec5}

\begin{figure}[t]
\centering
\includegraphics[width=0.7\textwidth]{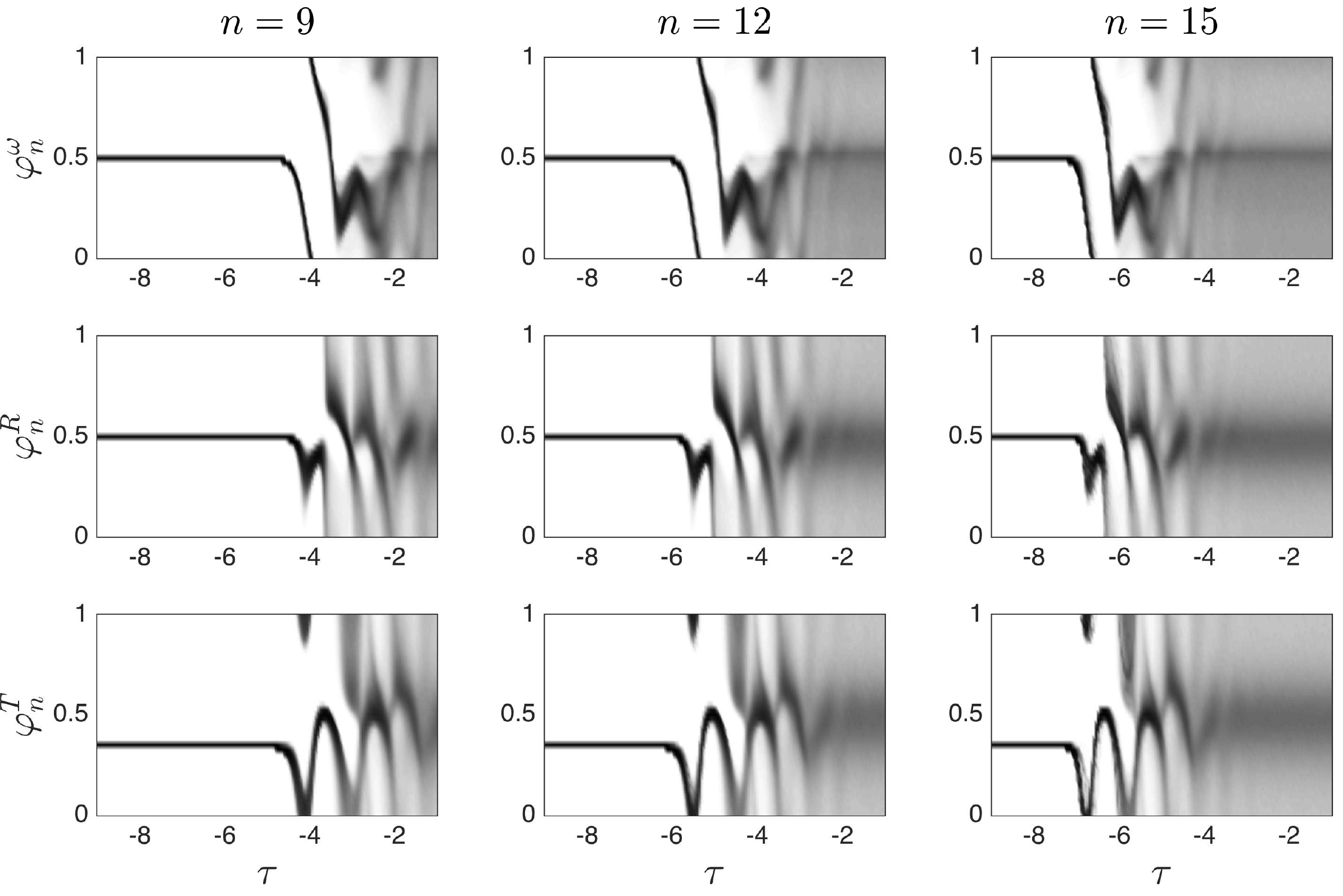}
\caption{Stochastic traveling wave of the RT instability in 2D model. Shown are PDFs of the variables $\varphi_n^{\omega}$, $\varphi_n^R$ and $\varphi_n^T$ that describe shell multipliers in (\ref{eq28}) and (\ref{eq28b}). PDFs are plotted using grayscale (darker color corresponds to a higher probability) as functions of renormalized time $\tau = \log_h t$ at shell numbers $n = 9,12,15$. The graphs at different shells are almost identical, with horizontal shifts due to wave speed $v = -2$.}
\label{fig3}
\end{figure}

Our theoretical construction in the previous section, which suggests that the RT instability represents a stochastic wave traveling with a constant speed in renormalized coordinates, is confirmed by numerical simulations in Fig.~\ref{fig3} for the 2D model. For statistical analysis $10^5$ independent simulations were used as specified in Section~\ref{sec3}. The figure shows probability density functions (PDFs) for the angles 
\begin{equation}
\varphi_n^{\omega} = \frac{1}{\pi}\mathrm{arctan}\, \rho_n^{\omega},\quad
\varphi_n^R = \frac{1}{\pi}\mathrm{arctan}\, \rho_n^{R},\quad
\varphi_n^T = \frac{1}{\pi}\mathrm{arctan}\, \rho_n^{T}.
\label{eq28b}
\end{equation}
The use of such variables allows a convenient representation of multipliers (\ref{eq28}) that accounts both for large and small values. Variables (\ref{eq28b}) have values in the interval $-1/2 \le \varphi_n \le 1/2$, where $\varphi_n = 0$ corresponds to $\rho_n= 0$ and $\varphi_n = \pm 1/2$ correspond to $\rho_n = \pm\infty$. This interval is extended periodically in the figure for better visualization. The traveling wave structure is seen very clearly connecting the two constant states at both sides. The constant state in front of the wave (at smaller times) corresponds to a deterministic state given by initial condition (\ref{eq14}), i.e., the PDFs are Dirac delta-functions. Behind the wave (at larger times), the constant state is stochastic with a continuous probability density independent of $\tau$. We will study this stationary stochastic state later in this section.

\begin{figure}[p]
\centering
\includegraphics[width=0.7\textwidth]{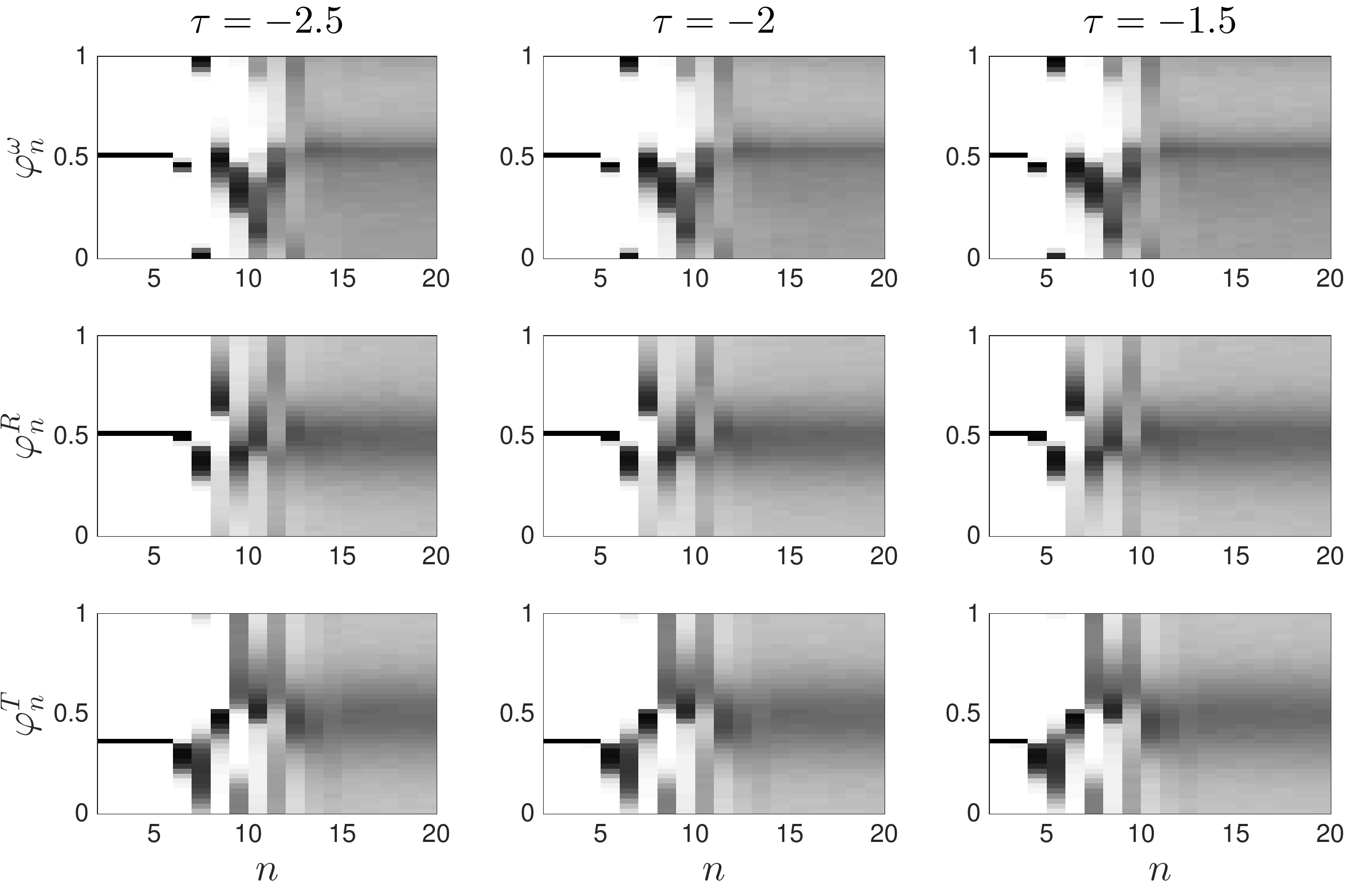}
\caption{Stochastic traveling wave of the RT instability in 2D model: same as in Fig.~\ref{fig3} but now represented as functions of shell numbers $n$ at different renormalized times $\tau = \log_h t$. The graphs at different times are almost identical, with the shifts by one and two shell numbers.}
\label{fig4}
\end{figure}

Representation of the same stochastic wave in also given in Fig.~\ref{fig4} from a different point of view. Here we fixed the renormalized time at $\tau = -2.5,\,-2,\,-1.5$ and showed PDFs depending on the discrete shell number $n$. Since the wave speed is $v = -2$, these times correspond to the traveling wave shifted exactly by one and two shells to the left, in full agreement with the numerical results shown in the figure. The wave has a finite spread in renormalized space-time: it extends roughly to $\Delta n \approx 6$ shells (Fig.~\ref{fig4}) and it passes over a given shell in time interval $\Delta\tau \approx \Delta n/|v| \approx 3$ (Fig.~\ref{fig3}). This corresponds to less than two decades of spatial scales $r_n$, featuring a strongly anisotropic and non-stationary transition to a developed turbulent state.

\subsection{Universal growth of the mixing layer}
\label{sec5.1}

Uniqueness of the chaotic attractor in the renormalized system provides a universal probability measure for the RT wave. In principle, multi-stability with several attractors is also possible, but it does not seem to be the case for our model. Uniqueness of the RT wave explains several important properties extensively studied both numerically and experimentally for the full Bussinesq approximation. The first property refers to the asymptotic growth of the mixing layer in (\ref{eq19}). It can be written as
\begin{equation}
L(t) = \alpha t^2 = \alpha h^{2\tau},
\label{eq29}
\end{equation}
where the coefficient $\alpha$ is expected to be universal (independent of initial conditions)~\cite{boffetta2016incompressible}. The corresponding 
\begin{equation}
N = -\log_h L = -a-2\tau,\quad a = \log_h \alpha,
\label{eq30}
\end{equation}
determines the smallest shell number involved in turbulent dynamics. Due to discrete nature of the shell model mentioned in the end of Section~\ref{sec3b}, we expect a periodic function $\alpha(\tau) = \alpha(\tau+1/2)$ instead of a constant coefficient, as confirmed in Fig.~\ref{fig2}.

With the representation of RT instability as a stochastic wave traveling with speed $v = -2$, relation (\ref{eq30}) follows directly. In this representation, the coefficient $a$ is arbitrary, because the renormalized system (\ref{eq24})--(\ref{eq26}) is translation invariant. However, the shift of a shell number by $-a$ corresponds to the multiplication by $h^{a} = \alpha$ of the original temperature variables $R_n$ and $T_n$ in (\ref{eq23}). Hence, only a specific choice of $\alpha$ matches the initial condition (\ref{eq14}) (a ``temperature jump'') with $T_n = 1$ in the deterministic state ahead of the wave. Thereby, universality of the coefficient $\alpha$ becomes a simple consequence of the uniqueness of the RT stochastic wave.

In continuous models~\cite{fermi1953taylor,cook2004mixing,ristorcelli2004rayleigh}, a transient due to a finite initial perturbation was taken into account by modifying (\ref{eq29}) as 
\begin{equation}
L(t) = \alpha (t-t_0)^2 = \alpha (h^\tau-t_0)^2,
\label{eq29b}
\end{equation}
where a small shift $t_0$ reflects the   translation invariance of the original system with respect to time. In this case (\ref{eq30}) becomes
\begin{equation}
N = -\log_h L = -a-2\log_h(h^\tau-t_0) = -a-2\tau+\frac{2t_0}{\log h}\,e^{-\tau\log h}+o(t_0),
\label{eq30b}
\end{equation}
where we used the Taylor expansion in $t_0$. The last expression shows that $t_0$ in (\ref{eq29b}) takes into account a perturbation of the RT stochastic wave described by an exponentially decaying mode with the eigenvalue $\lambda_1 = -\log h \approx -0.69$.

\begin{figure}
\centering
\includegraphics[width=0.8\textwidth]{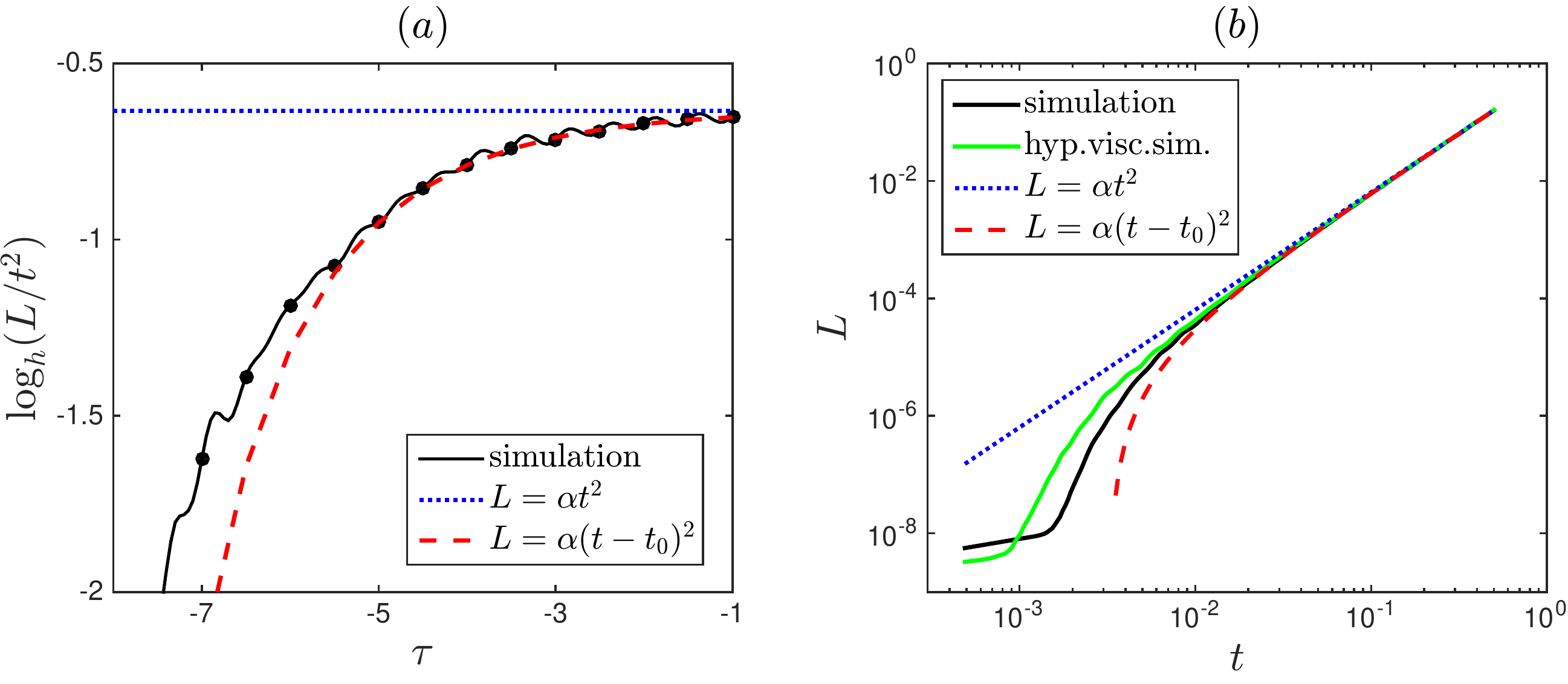}
\caption{Evolution of the width $L$ of the mixing layer with time $t = h^\tau$: (a) for $\log_h(L/t^2)$ and (b) for $L$ (log-scale). Results of numerical simulations (black) are compared with the asymptotic relations (\ref{eq29}) and (\ref{eq29b}) (dotted blue and dashed red). Black dots on the left panel correspond to discrete times $\tau = \ldots,-2,-1.5,-1$. A green curve on the right panel corresponds to simulations with hyper-viscosity.}
\label{fig5}
\end{figure}

Figure~\ref{fig5}(a) presents the time dependence of $\log_h(L/t^2)$, where black line corresponds to the numerical simulations and the red line to the least-squares fit with expression (\ref{eq29b}). The fitting is made in the interval $-5 \le \tau \le -1$ and yields the parameters 
$\alpha = 0.644$ and $t_0 = 0.0032$.
Note that there is a technicality related to the discreteness of spatial scales in the shell model, which is responsible for a small periodic oscillation around the mean curve. For numerical fits, it is convenient to consider the associated discrete (half-integer for $v = -2$) times $\tau = \ldots,-2,-1.5,-1$, when the RT wave moves by an integer number of shells.
At times $\tau \gtrsim -5$, the agreement in Fig.~\ref{fig5}(a) is very accurate, which suggests that $\lambda_1$ is the largest eigenvalue for the stationary RT wave. Deviations are large at earlier times, closer to the moment when the RT wave is initialized.

To confirm the universality of the coefficient $\alpha$ numerically, we performed a similar series of $10^5$ simulations, but with a different (super-viscous) dissipation mechanism: $k_n^2$ was substituted by $k_n^3$ in all dissipation terms of model (\ref{eq10})--(\ref{eq12}). For the dissipation parameters we used $\nu = \kappa = 10^{-23}$. The results are shown in Fig.~\ref{fig5}(b) by a green line, converging to the asymptotic expression with the same value of $\alpha$. Also, these simulations provide the same form of a stochastic RT wave as observed in Figs.~\ref{fig3} and \ref{fig4}.

\subsection{Intermittency and (an)isotropy at small scales}
\label{sec5.2}

\begin{figure}
\centering
\includegraphics[width=0.8\textwidth]{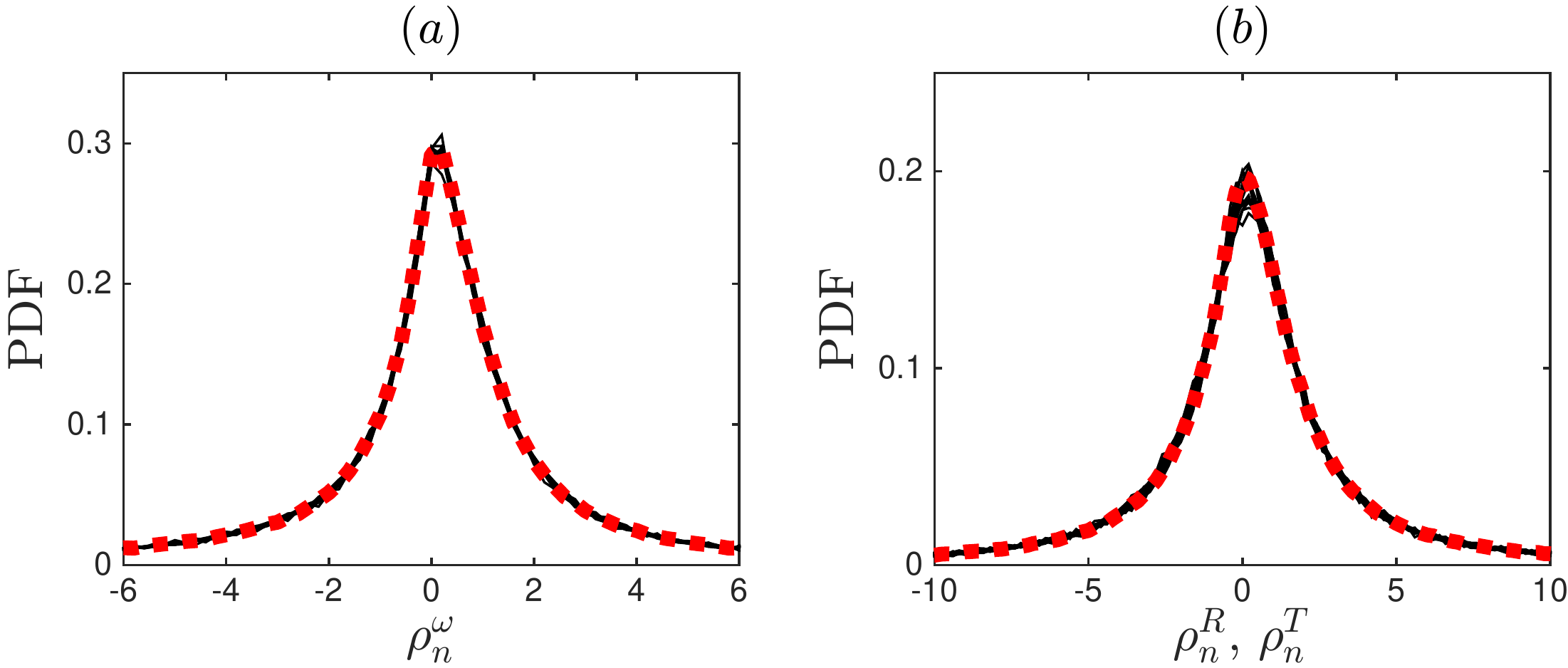}
\caption{PDFs for (a) $\rho_n^{\omega}$ and (b) $\rho_n^R$ and $\rho_n^T$ at the time $\tau = -1.5$. Shown are a number of black solid curves (all collapsed to a single profile), which correspond to the shells $n = 13,14,\ldots,21$. The red dotted lines correspond to the stationary turbulent state (for the shell $14$ in the inertial range) obtained by a single large-time numerical simulation with parameters $\nu = \kappa = 10^{-13}$ and constant forcing at the first shell.}
\label{fig6}
\end{figure}

The constant state behind the RT wave corresponds to small scales, and it is clearly seen as grey time- and scale-independent regions on the right side in Figs.~\ref{fig3} and \ref{fig4}. The fact that the dynamics at these scales is not any more influenced by initial conditions implies that the corresponding statistics represents a stationary turbulent state. Recall that the third Kolmogorov hypothesis, which we employed for the definition of multipliers, suggests precisely that the statistics of multipliers is universal and scale independent for the stationary turbulence. This fact is persuasively
confirmed by numerical simulations in Fig.~\ref{fig6}. Here several thin black lines are shown that correspond to PDFs for the shells $n = 13,14,\ldots,21$, and these lines collapse perfectly to a single curve. The time $\tau = -1.5$ chosen in this figure is the same as for the right panels of Fig.~\ref{fig4}. These results match precisely with the dotted red lines in Fig.~\ref{fig6} representing the PDFs in the inertial range for the stationary turbulent state. The latter is obtained by a single large-time numerical simulation of model (\ref{eq10})--(\ref{eq12}) with constant forcing at the first shells. It is remarkable that the PDFs for the horizontal and vertical temperature fluctuations, shown together in Fig.~\ref{fig6}(b) are identical. This means that the RT instability recovers isotropy at small scales for the 2D shell model.

\begin{figure}
\centering
\includegraphics[width=0.95\textwidth]{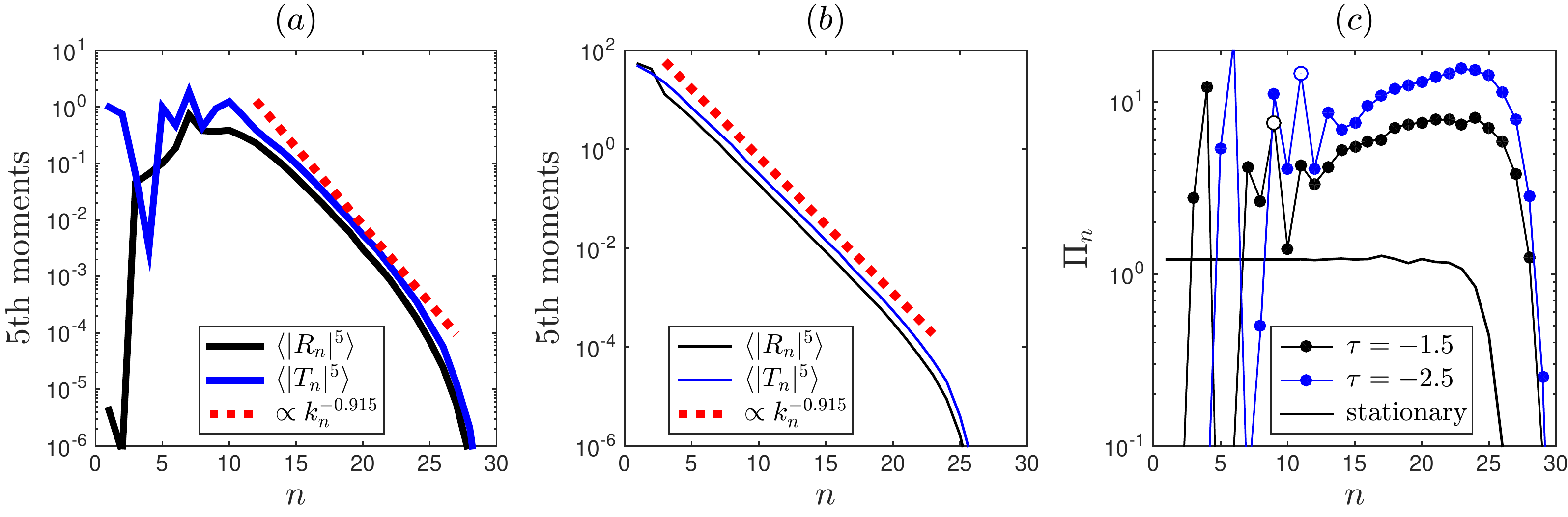}
\caption{(a) Moments $\langle |R_n|^5\rangle$ and $\langle |T_n|^5\rangle$ at late time $\tau = -1$, averaged with respect to ensemble of $10^5$ simulations. (b) Same moments but for the developed turbulent state, averaged with respect to time. (c) Flux of entropy at different times for the RT instability (averaged with respect to ensemble of $10^5$ simulations at times $\tau = -1.5$ and $-2.5$) and for the stationary turbulence (averaged with respect to time). White dots indicate negative flux values.}
\label{fig7}
\end{figure}

Universal multipliers imply, in general, anomalous scaling for the moments of shell variables in stationary turbulence~\cite{eyink2003gibbsian}.  Figure~\ref{fig7} shows the fifth moments of temperature variables, $\langle|R_n|^5\rangle$ and $\langle|T_n|^5\rangle$ at $\tau = -1$ in panel (a), which are compared with the results obtained by time-averaging for the developed turbulent state shown in panel (b). From the latter, one can see that the moments develop power laws with the anomalous exponent, $\propto k_n^{-0.915}$, which deviates from the dimensional prediction $k_n^{-1}$ following from (\ref{eq21}). The same power-law slope is indicated in Figure~\ref{fig7}(a) showing only a weak agreement with the stationary state. The two important conclusions can be drawn from these results. First, the moments $\langle |R_n|^5\rangle$ and $\langle |T_n|^5\rangle$ in both figures are not exactly the same, having a small but persistent vertical shift. This means that the isotropy clearly observed for the multipliers in Fig.~\ref{fig6}(b) does not extend to the original temperature variables. Second, the RT instability develops power-laws for the variables $R_n$ and $T_n$ quite poorly in drastic contrast to the almost perfect power-laws for multipliers in Fig.~\ref{fig6}. These conclusions support the third Kolmogorov hypothesis suggesting the multipliers as appropriate variables for the turbulence description. 

\begin{figure}
\centering
\includegraphics[width=0.65\textwidth]{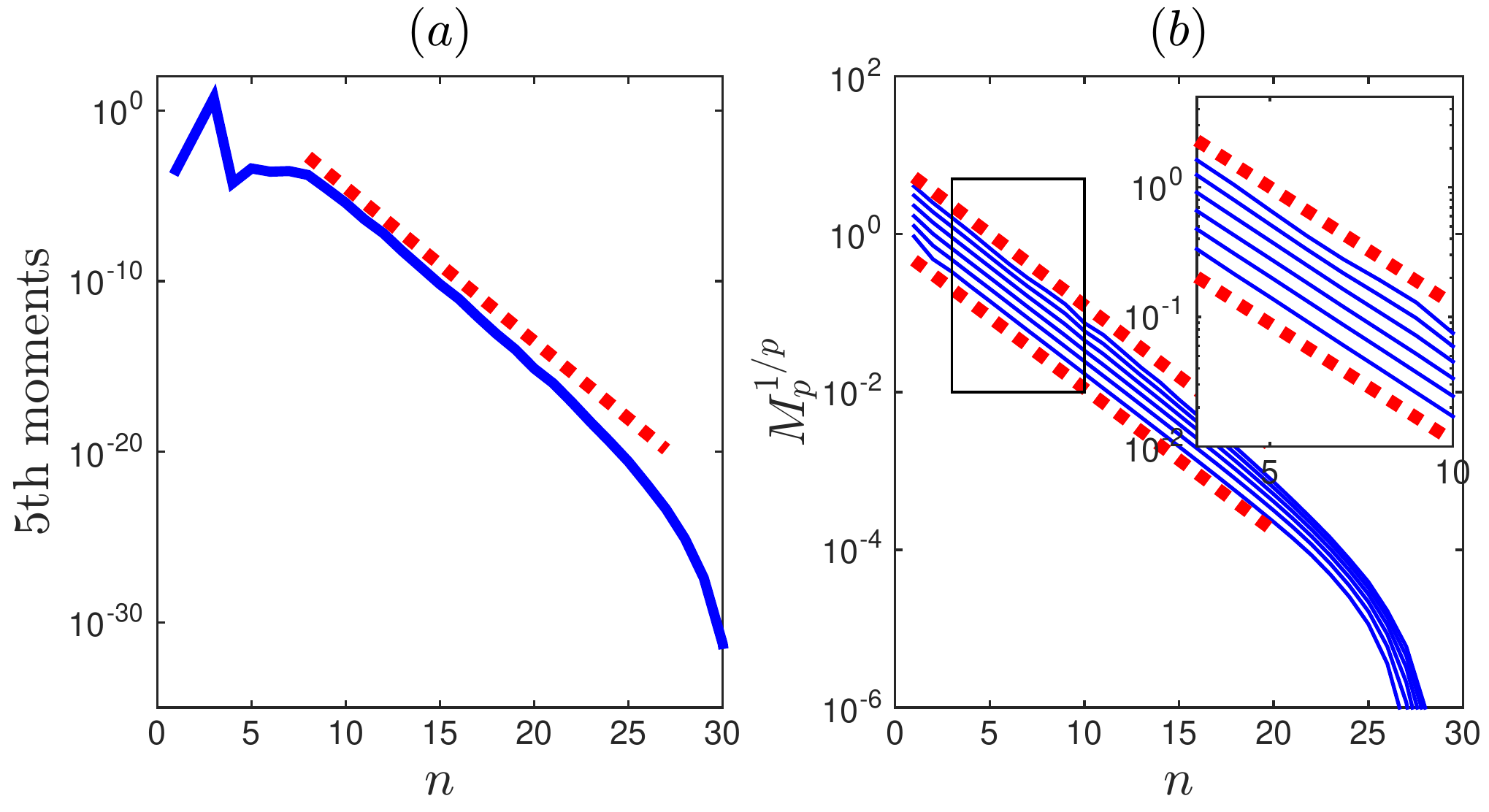}
\caption{(a) Fifth moments of shell velocities $\langle|u_n|^5\rangle$ at late time $\tau = -1$, averaged with respect to ensemble of $10^5$ simulations. (b) Weighted moments of shell velocities $M_p^{1/p}$ for the developed turbulent state, averaged with respect to time; results are shown for $p = 1,2,\ldots,6$ (higher moments correspond to higher solid curves). The dotted red lines show the dimensional prediction for the slope $M_p^{1/p} \propto k_n^{-3/5}$}.
\label{fig_new2}
\end{figure}

Statistical analysis suggests that shell velocities are not intermittent. In Fig.~\ref{fig_new2}(a) the fifth moment is shown at the same time as in Fig.~\ref{fig7}(a), with the dotted red line representing the dimensional scaling (\ref{eq21}). Though the comparison is not convincing, one may consider the analogous results for the developed (forced) turbulence. Using the same simulation as in Fig.~\ref{fig7}(b), we compute the velocity moments $M_p = \langle|u_n|^p\rangle$ averaged with respect to time. The weighted moments $M_p^{1/p}$ are shown in Fig.~\ref{fig_new2}(b) for $p = 1,2,\ldots,6$. The inertial interval gets shorter for larger moments, but one can clearly see that the statistics of velocities is not intermittent within the numerical accuracy (enlarged initial part of the inertial interval is presented in the inset).

Further insight on the difference between the original variables and multipliers can be obtained by looking at the entropy $S = \sum(R_n^2+T_n^2)$, which is an inviscid invariant for our shell model. The corresponding entropy flux from shell $n$ to shell $n+1$ is given by
\begin{equation}
\Pi_n = -2 \omega_n (T_nT_{n+1}+R_nR_{n+1}).
\label{eq35}
\end{equation}
In the phenomenological theory of RT instability (Section~\ref{sec2}) a mean value $\varepsilon_T(t) = \langle \Pi_n \rangle$, averaged over a statistical ensemble at given time, is assumed to be scale-independent in the inertial interval. This means that small-scale dynamics is assumed to be quasi-stationary, dominated by a slowly changing mean dissipation rate $\varepsilon_T(t)$. The fluxes averaged over the ensemble of $10^5$ simulations at fixed times $\tau = -1.5$ and $\tau = -2.5$ are compared in Fig.~\ref{fig7}(c) with the time-averaged values for the developed turbulent state. We see that the assumption of quasi-stationarity at small scales is rather poorly satisfied for the RT instability despite a large extent of the inertial range. This once again shows a drastic difference of the weak convergence for original variables, as opposed to the very fast and accurate convergence for  multipliers at small scales, indicating the latter as proper variables for description of turbulence. Note that oscillations in Fig.~\ref{fig7}(c) correspond to analogous oscillations in the front part of the stochastic wave in Fig.~\ref{fig4}.

\section{RT instability in 3D case}
\label{sec6}

The 3D RT instability is modeled by choosing the parameter $c = 1/h^2$ in (\ref{eq13}), in which case the nonlinear term in vorticity equation (\ref{eq10}) conserves the kinetic energy $E = \sum u_n^2$.
In this section we demonstrate that the RT instability in our model is similar for the 3D and 2D cases, in the sense that both are described by a stochastic wave traveling from small to large scales. We choose the coupling parameter in Eqs.~(\ref{eq11}) and (\ref{eq12}) as $\gamma = 0.7$ for numerical simulations, which leads to chaotic behavior for the RT instability. With the parameter $\gamma = 1$, used earlier in the 2D model, the dynamics becomes regular (quasi-periodic as in~\cite{mailybaev2016spontaneous}), making this choice less attractive for our purpose. 

Recall that the phenomenological theory summarized in Section~\ref{sec2} predicts that the energy cascade to small scales dominates the statistics in the inertial interval, while the buoyancy becomes a passively advected scalar. In this approximation, we can find a stationary solution in our 3D shell model at small scales. To find this solution explicitly, it is convenient to introduce a complex variable $\theta_n = R_n+iT_n$ for the total temperature fluctuation at scale $r_n$.  Then, equating the right-hand sides in (\ref{eq10})--(\ref{eq12}) to zero and neglecting the buoyancy term $k_nR_n$, yields
\begin{equation}
\omega_n = k_n u_n = \alpha_1 k_n^{2/3},\quad
\theta_n = \alpha_2 \zeta^n+\alpha_3{\overline{\zeta}}^{\,n},\quad
\zeta = \frac{i\gamma}{2} \pm \sqrt{-\frac{\gamma^2}{4}+h^{-2/3}},
\label{eqF1}
\end{equation}
where $\alpha_1 \in \mathbb{R}$ and $\alpha_2,\alpha_3\in\mathbb{C}$ are arbitrary factors. 
The scaling of (\ref{eqF1}) agrees exactly with the dimensional prediction (\ref{eq20}), because $|\zeta| = h^{-1/3}$.
Note that the solution with shell velocities $u_n = \alpha_1 k_n^{-1/3}$ in (\ref{eqF1}) can be interpreted as a shock wave~\cite{mailybaev2015continuous}. Also, such scaling of shell velocities yields a constant energy flux from large to small scales.

\begin{figure}[t]
\centering
\includegraphics[width=0.8\textwidth]{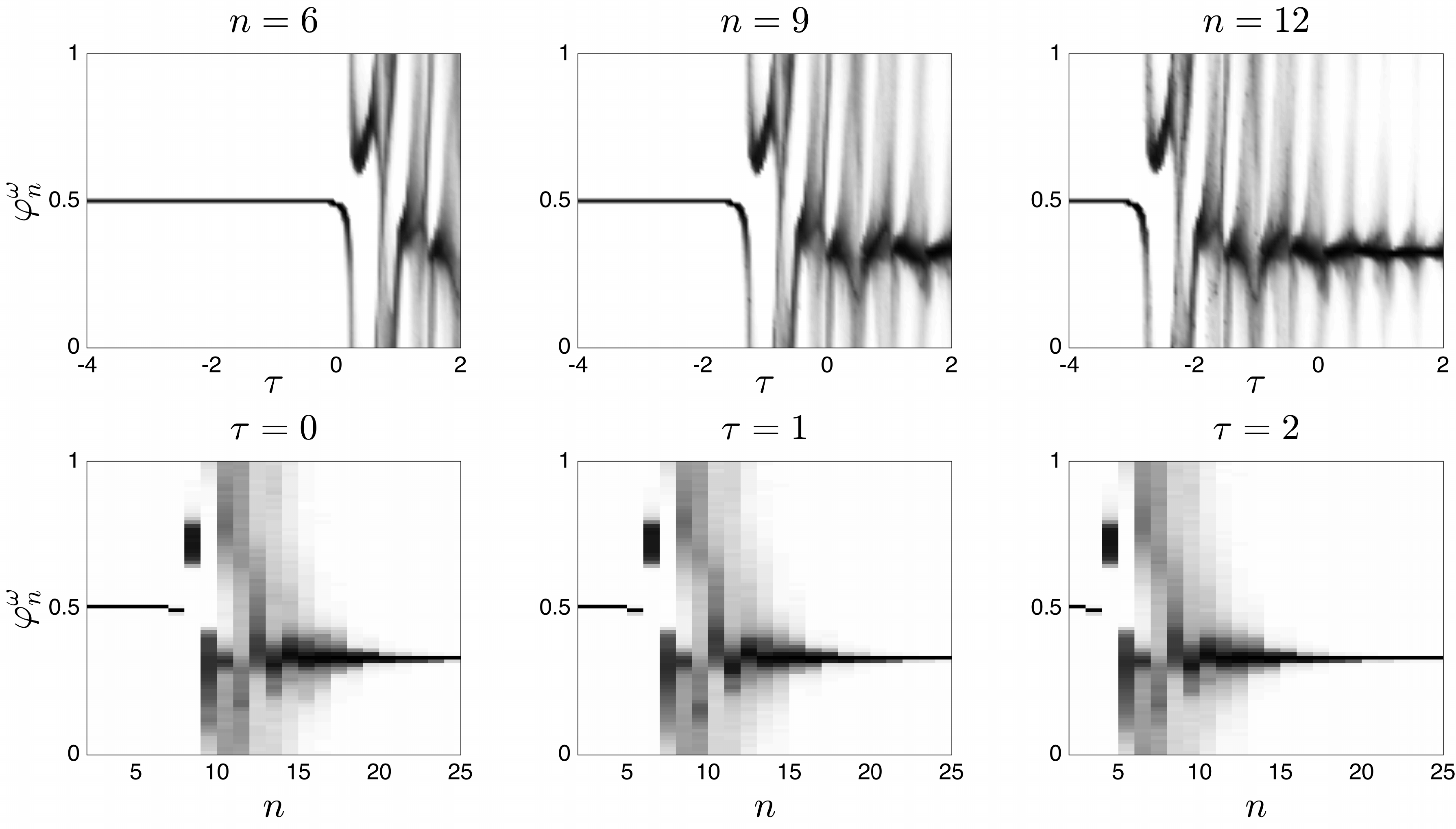}
\caption{Stochastic traveling wave of the RT instability for 3D model. Shown are the PDFs of variables $\varphi_n^{\omega}$ that describe the multipliers for vorticity in (\ref{eq28}) and (\ref{eq28b}). PDFs are plotted using grayscale (darker color corresponds to a higher probability) as functions of renormalized time $\tau = \log_h t$ at shell numbers $n = 6,9,12$ (first row), and as functions of shell numbers $n$ at different renormalized times $\tau = \log_h t = 0,1,2$ (second row). The graphs at different shells are almost identical, shifted according to the wave speed $v = -2$.}
\label{fig8}
\end{figure}

For the RT instability, we observe a stochastic wave traveling with constant speed in renormalized coordinates, in full agreement with the theory of Section~\ref{sec4}. Numerical evidence of this fact is demonstrated in Fig.~\ref{fig8},  presenting the results for multipliers of vorticity variables. The stochastic RT wave is seen very clearly as almost identical PDF patterns shifted horizontally according to the wave speed $v = -2$. At small scales (large $n$) the solution tends to a constant deterministic solution (\ref{eqF1}): the PDF represents a Dirac delta-function at $\rho_n^{\omega} = \omega_n/\omega_{n-1} = h^{2/3}$, see second row in Fig.~\ref{fig8}. For the temperature variables, simulations yield a similar behavior apart from a more sophisticated asymptotic at small scales, which we expect to agree with (\ref{eqF1}).
Following the argument of Section~\ref{sec5.1}, we associate the stochastic RT wave with a chaotic attractor, which explains the universal quadratic growth (\ref{eq29}) of the mixing layer in Fig.~\ref{fig2}(b).

Despite our shell model does not quite reproduce a typical behavior of the 3D Boussinesq system, where the dynamics at small scales is chaotic, it brings an important message that distinguishes the RT instability from the developed turbulence. In our case, the ``developed turbulent state'' given by (\ref{eqF1}) is regular, while the RT instability is intrinsically stochastic. Therefore, the stochastic component in our example is an attribute of the RT wave only, which has deterministic constant states at both large- and small-scale sides, Fig.~\ref{fig8}. A range of scales occupied by the RT wave extends to almost 10 shells or, equivalently, to almost three decades of scales $r_n = h^{-n}$. 

\section{Conclusions}

In this paper we argue that turbulent development of the Rayleigh--Taylor (RT) instability (the instability of an interface between fluids of different density under the action of gravity) can be described as a stochastic traveling wave in a renormalized system. A proposed renormalization scheme uses logarithmic time and space variables, and the RT wave is associated with a probability measure of a chaotic attractor in a usual dynamical system sense. The infinite-time dynamics in this setting is induced by mapping the initial time $t = 0$ to the renormalized time $\tau = \log_h t \to -\infty$. Furthermore, with the third Kolmogorov hypothesis suggesting that the turbulence can be described using ratios of velocity increments (multipliers), we arrive to a simple picture of a steady-state wave traveling with a constant speed between two constant limiting states. These constant states correspond to a deterministic initial condition (temperature jump) at large scales and to developed turbulence at small scales. It is shown how the existence of such a wave leads to various universal properties of the RT instability, e.g. the universal growth of a mixing layer and scaling laws. 

The analysis is performed using a new shell model that is designed to feature basic properties of the phenomenological theory for the RT instability. This is done both for the two- and three-dimensional cases. Following theoretical arguments, we perform $10^5$ independent numerical simulations that persuasively confirm the predicted form of a stochastic solution. Also, numerical simulations verify several properties of the RT instability that are hard to access accurately in full convection models. We show that the RT instability in the 2D case recovers isotropy at small scales in terms of the multipliers, but not in original variables. Furthermore, the multipliers demonstrate a very fast and accurate convergence to universal distributions at small scales, while this is again not the case for original variables. In the 2D model, intermittency is demonstrated for temperature variables, while velocities appear to be not intermittent. The results provide a traveling wave that occupies an interval of scales up to two decades for the 2D shell model and three decades in the 3D shell model. This means that a similar structure may be accessible (especially for the 2D case) in the full continuous model with modern computational resources. 

Our results provide the new terminology that goes beyond the phenomenological and dimensional theories. Namely, a representation that maps the solution into the stochastic wave may help for understanding the full mechanism of the RT instability. In a more general sense, this approach explains spontaneously stochastic solutions emerging from singular initial conditions in the inviscid limit. Here the traveling wave representation serves to justify the uniqueness (universality) of the resulting spontaneously stochastic process. 
It can be expected that a similar mechanism underlines other turbulent phenomena initiated by singular initial conditions like, e.g., the Kelvin--Helmholtz instability.

\section*{Acknowledgments} 
This work was supported by the 
CNPq (grant 302351/2015-9) and the Program FAPERJ Pensa Rio (grant E-26/210.874/2014).

\bibliographystyle{plain}
\bibliography{refs}

\end{document}